\documentclass[aps,prl,twocolumn,floatfix,superscriptaddress]{revtex4-1}
\usepackage{graphicx}
\usepackage{color}
\usepackage{graphicx}
\usepackage{amssymb}
\usepackage{amsthm}
\usepackage{amsmath}
\begin{document}
%==============================================================================
\title{Solid-liquid interfacial premelting}

\author{Yang Yang}
\affiliation{Department of Chemistry, University of Kansas, Lawrence,
KS 66045, USA}
\author{Mark Asta}
\affiliation{Department of Materials Science and Engineering, University of
California, Berkeley, CA 94720, USA}
\author{Brian B. Laird}
\thanks{Author to whom correspondence should be addressed}
\affiliation{Department of Chemistry, University of Kansas, Lawrence,
KS 66045, USA}
\date{\today}
%==============================================================================
\begin{abstract}
 We report the observation of a premelting transition at chemically sharp solid-liquid interfaces using molecular-dynamics simulations.  The transition is observed in the solid-Al/liquid-Pb system and involves the formation of a liquid interfacial film of Al with a width that grows logarithmically as the bulk melting temperature is approached from below, consistent with current theories of premelting.  The premelting behavior leads to a sharp change in the temperature dependence of the diffusion coefficient in the interfacial region, and could have important consequences for phenomena such as particle coalescence and shape equilibration, which are governed by interfacial kinetic processes.
\end{abstract}
\maketitle

The term premelting refers to the formation of thermodynamically stable liquid films at solid interfaces subjected to temperatures below but near the bulk melting temperature ($T_m$)~\cite{Chernov93,Rosenberg05}.  Surface premelting (SP), the formation of a premelting layer at a solid-vapor interface, was first observed experimentally by Frenken and van der Veen~\cite{Frenken85} using proton scattering. Since then, advances in experimental techniques have provided powerful tools for direct atomic-resolution observations of this surface phase transition~\cite{Prince88,Yang89,Feenstra91, Blanckenhagen94, Theis95, Wang98,Li07} These experimental studies have been complemented by a number of detailed atomistic simulations probing the existence and atomic-level mechanisms of SP (Refs.~\onlinecite{Dash06,Matsui91,Gravil96,Tartaglino05,Song10,Bavli11,Williams09,Mishin10} and references therein).  Premelting at solid-solid interfaces has also been reported in the literature and can take two basic forms: premelting at solid-solid heterophase boundaries and grain-boundary premelting (GBP). Examples of the former are found at interfaces between Pb and Al\cite{Dahmen04}, SiO and Al$_2$O$_3$\cite{Kofman94}, as well as in ice at solid substrates, which plays a role in frost heave \cite{Dash06,Rempel01,Hansen-Goos10,Rempel04}. Premelting at grain boundaries has been the subject of numerous continuum~\cite{Lobkovsky02,Tang06a,Tang06b, Berry08,Mellenthin08,Mishin09b,Cogswell11}, atomistic (Ref.~\onlinecite{Mishin10} and references therein) and experimental\cite{Luo07} studies.

In this work, we report molecular-dynamics (MD) simulation results on Al/Pb solid-liquid interfaces that predict the existence of a third class of premelting transition, namely, solid-liquid premelting (SLP).  In this process a premelting liquid layer forms at a solid-liquid interface, below $T_m$ of the solid. Such a transition should, in principle, be possible at chemically heterogeneous solid-liquid interfaces in which the melt phase of the solid  and the liquid phase are mutually immiscible, as is the case with Al and Pb near $T_m$ for Al.   Solid-liquid premelting has not been previously reported under ambient pressures, either experimentally or by simulation - although some evidence of SLP have been reported in simulations of solid-liquid interfaces under extremely high pressure (diamond anvil) \cite{Sorkin06,Starikov09}. As we will demonstrate, the process of SLP leads to a sharp change in the temperature dependence of the diffusion coefficient in the interfacial region, which is expected to have important consequences for the kinetics of interface-controlled processes such as particle shape equilibration or coalescence of liquid nanoparticles  in solids governed by interface-mediated Brownian motion \cite{Johnson04a,Johnson04b}.

Thermodynamically, premelting occurs near $T_m$ when the interfacial free energy, $\gamma_{\mathrm{s}\alpha}$, between the solid and another phase $\alpha$  ($\alpha$ = solid, liquid or gas) is larger than the sum of that for the solid-melt ($\gamma_{\mathrm{sl}}$) and melt-$\alpha$ ($\gamma_{\mathrm{l}\alpha}$) interfaces:
\begin{equation}
\Delta \gamma = \gamma_{\mathrm{s}\alpha}  - [\gamma_\mathrm{sl}+\gamma_{\mathrm{l}\alpha}]  > 0
\label{eq:delgam}
\end{equation}
Thus, if the undercooling ($\Delta T = T_m - T$) is not too great, it is thermodynamically favorable to form a thin film of metastable liquid because the increase in bulk free energy is more than compensated by a lowering of the total interfacial free energy. The width of the interface as a function of undercooling depends on $\Delta \gamma$ and $\Delta T$, as well as the nature of the potential of interaction between the two interfaces - the so-called "disjoining potential". The pioneering theoretical studies by Kikuchi and Cahn~\cite{Kikuchi80} on grain-boundary premelting and Lipowsky~\cite{Lipowsky82} on SP both predict a logarithmic dependence\cite{disclaimer} of the width of the premelting layer with respect to the undercooling~\cite{Lipowsky83}:
\begin{equation}
w(\Delta T)= -w_0 \ln{[\Delta T/T_0]}
\label{eq:wlog}
\end{equation}
where $w_0$ and $T_0$ are constants specific to the given interface.

{\em The System:} The Al-Pb system is an ideal model alloy for the study of chemically heterogenous solid-liquid  interfaces. The phase diagram is a simple monotectic that has a broad liquid-liquid miscibility gap, negligible solubility of Pb in the Al solid phase and a large melting point separation (600 K for Pb and 933 K for Al). We have previously reported results from  MD  simulations on this system at 625 K, a temperature just above the melting point of Pb.~\cite{Yang12} The simulation results are consistent with an \emph{in situ} transmission-electron-microscopy (TEM) study of liquid Pb inclusions embedded in a crystalline Al matrix~\cite{Gabrisch01} in that the (111) interface is shown to be faceted, while (110) and (100) are rough at this temperature. The experiments also show that the (111) interface undergoes a roughening transition about 100 K below the melting point of Al.

{\em Simulation Details:} In our simulations of the Al/Pb solid-liquid interface, we employ a classical many-body potential developed by Landa, {\em et al.}~\cite{Landa00} to model the interatomic interactions. This potential predicts mutual immiscibility of Pb and Al in the solid state and a large-liquid state miscibility gap consistent with the experimental phase diagram up to 1200K. The melting points of this potential~\cite{Yang12} are 615.2(2)K and 922.4(2)K for Pb and Al, respectively. The MD simulations are performed using LAMMPS~\cite{Plimpton95}. Equilibrated SL interfaces are set up at varying temperatures, $T$, ranging from 625 to 900K, separated by 25K up to 90K with additional simulations at  912, 920, 921 and 922K. Three crystallographic orientations for the Al-Pb interfaces are examined: (100), (110) and (111).

To produce the equilibrated interfaces, constant-area, constant-normal-pressure MD ($NP_zAT$) simulations up to 50 ns in length are used  to yield the appropriate equilibrium number density, $\rho$, pressure (1 bar) and composition. These are followed by constant $NVT$ simulations to collect production data. Five replica systems (each containing two independent interfaces) are used at each temperature and orientation to improve statistics. For additional details as to the methods of interface set-up, equilibration and analysis, see  the supplemental materials\cite{SM} and Ref.~\onlinecite{Yang12}.

{\em Characterization:} The solid-liquid interface is characterized through the determination of interfacial profiles, which show the change in specific properties (e.g., density, local structural order, composition and diffusion constant) as functions of the distance normal to the interfacial plane, defined as the $z$ direction.  The $z$-coordinate is measured relative to a Gibbs Dividing Surface (GDS), defined here such that the excess number of Al atoms, $\Gamma_\mathrm{Al}$ is zero.~\cite{SM} To determine the extent of the premelting layer, we utilize two  different order-parameter (OP) profiles. The first profile uses a local structure OP that distinguishes solid from liquid phases~\cite{Morris02} and is normalized to be 1 in the solid phase and 0 in the liquid phase. The second is a compositional order parameter equal to 1 in a pure Al system and 0 for pure Pb. For a Al/Pb solid-liquid interface without a premelting layer, these two order-parameter profiles will be approximately coincident; however, in the presence of a premelting layer (a liquid Al layer separating solid Al from liquid Pb), the interfacial position indicated by these two OP profiles will be separated by the width of the premelting layer, $w$.

%~~~~~~~~~~~~~~~
\begin{figure}[h]
\includegraphics[width= .90\columnwidth]{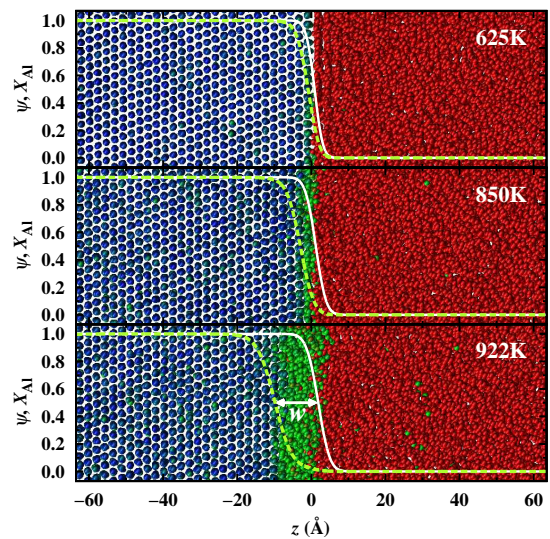}
\caption{ Snapshots of Al-Pb (111) solid-lilquid interfaces at three different temperatures. Top panel: 625K,  interface is faceted~\cite{Yang12}; Middle panel: 850K, interface is rough; Bottom panel: 922K, interface is premelted. Al atoms are colored coded on the structural OP as solid (blue) and premelted liquid (green). Pb atoms are shown as red. In each image, the average structural OP and chemical OP are plotted as dotted green and solid white lines, respectively. The premelting width $w$ is defined as the distance between the two OPs (at half value).}
\label{fig1}
\end{figure}
%~~~~~~~~~~~~~~~

Figure~\ref{fig1} shows $NVT$ snapshots of equilibrated (111) Al-Pb interfaces at increasing temperature for three different temperatures (625, 850 and 922K), together with the corresponding time-averaged structural and composition OP profiles. For low $T$ (top panel) just above $T_m^\mathrm{Pb}$, the (111) interface is faceted~\cite{Yang12} and two profiles are nearly coincident. For high $T$ (bottom panel) just below $T_m^\mathrm{Al}$, the two profiles are separated by nearly 10$ \mathrm{\AA}$ indicating the presence of a premelting layer of liquid Al sandwiched  between solid Al and liquid Pb.   In the central panel at 850 K, only the first complete layer of Al at the interface is structurally disordered.

\begin{figure}[h]
\includegraphics[width= 0.9\columnwidth]{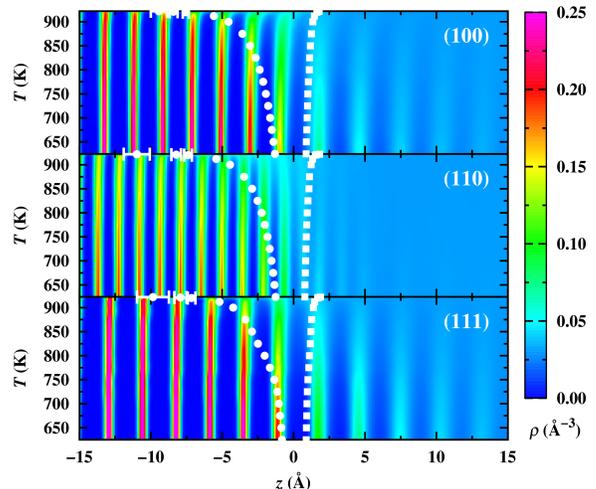}
\caption{ Equilibrium total number density contour maps of (100), (110) and (111) Al-Pb solid-liquid interface. Interface positions determined from structural OP profiles (circle) and chemical OP profiles (square) as function of $T$ are plotted on top of the correspondingly contour map. The error bars represent 95\% confidence levels. For symbols without error bars, the error is smaller than the size of the symbol.}
\label{fig2}
\end{figure}

{\em Results:} Color-contour plots of the  fine-scale density profiles, $\rho(z,T)$, are shown in Fig. 2 for the (100) (110) and (111) Al-Pb solid-liquid interfaces as a function of temperature and distance ($z$) normal to the interface. Also, plotted in Fig.~\ref{fig2} are the temperature-dependent interfacial positions defined by the midpoints of the structural and compositional order parameters. The distance between these two interfacial positions diverges as the Al melting point is approached, due to the formation of the premelting layer. The peaks of the density profiles are seen as vertical striations in the plot, which are stronger and more highly localized in the solid phase - to the left of the structural order parameter. To the right of the structural order parameter line, the density peaks are smaller in magnitude and more diffuse, consistent with the usual structural ordering of a liquid near a surface.  The much smaller magnitude of the liquid structural ordering in (110) - relative to (100) and (111) - was previously noted and discussed in Ref.~\onlinecite{Yang12}. Note that the position of the interface as defined by the composition OP is roughly independent of temperature, due to the mutual imiscibility of liquid Pb in both solid and premelted Al. The slight shift of this position towards higher $z$ near $T_m$ for Al is due to the lower density (relative to the solid) of the growing premelted Al layer.

Figure~\ref{fig2} shows that solid-liquid premelting occurs in this system for all orientations studied, with nearly identical behavior. This is in contrast to surface premelting in many fcc metal surfaces, in which (110) surfaces are prone to premelting, while premelting is not seen in the other orientations~\cite{Blanckenhagen94,Tartaglino05}.

%~~~~~~~~~~~~~~~
%~~~~~~~~~~~~~~~

%~~~~~~~~~~~~~~~
\begin{figure}[h]
\includegraphics[width= 0.9\columnwidth]{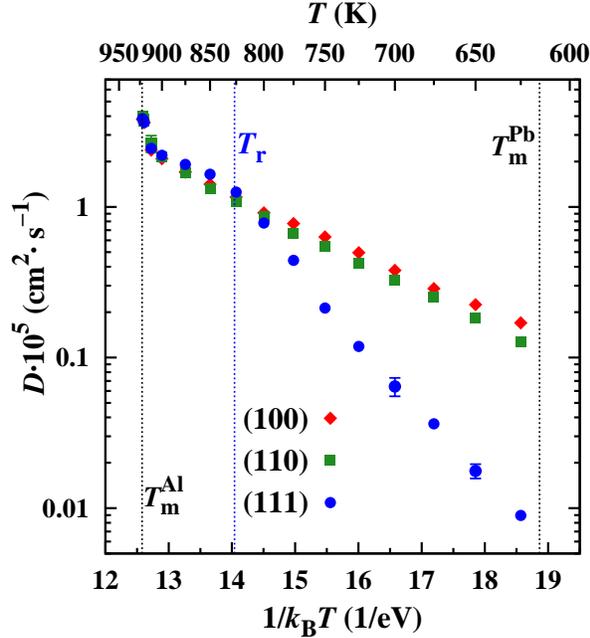}
\caption{Log-linear plot of the diffusion constant for Al atoms in the first Al layer as a function of $1/k_\mathrm{B} T$ for the (100), (110) and (111) -  diamonds, squares and circles, respectively. The Al melting point and estimated roughening temperature, $T_\mathrm{r}$, are indicated by vertical lines. The scale at the top of the figure shows the temperatures corresponding to the inverse temperatures shown on the independent axis.}
\label{fig3}
\end{figure}
%~~~~~~~~~~~~~~~

Our previous simulations at 625K\cite{Yang12} suggested that the (100) and (110) interfaces are rough at that temperature, while the (111) interface is faceted. However, visual inspection (together with orientational order parameter analysis) indicates that the (111) interface undergoes a roughening transition at higher temperatures (for example, see the middle panel of Fig.~\ref{fig1}). To probe this behavior in further detail, we examine particle transport in the first layer of Al., which is characterized through the calculation of diffusion-constant profiles, $D(z)$, determined from the mean-squared particle displacement versus time - see Ref.~\onlinecite{Yang12} for details. The full diffusion constant analysis is presented in the Supplemental Information,\cite{SM} but here we focus on the diffusion constant measured within the first Al layer (that is, the particles making up the density peak closest to $z=0$ on the negative side in Fig.~\ref{fig2}). Figure~\ref{fig3} shows a log-linear plot of diffusion constant versus $1/k_\mathrm{B}T$ for the particles in the first Al layer for the three interfacial orientations studied. The magnitude of the slope of this Arrhenius plot of $D$ can be interpreted as an activation energy for diffusion. Except very close to $T_\mathrm{m}^\mathrm{Al}$, the slope for the rough (100) and (110) interfaces is constant. For (111), however, the slope undergoes a discontinuous change at a temperature of about 826(4)K, indicating a sudden decrease in the activation energy for diffusion in this layer to a value that is comparable to that of the rough (100) and (110) interfaces. This temperature where the change in activation energy occurs roughly where the width of the interface begins to show a logarithmic dependence on undercooling (see below).  The temperature is also very near the roughening temperature of 823 K previously reported for the (111) interface based on {\em in situ} TEM experiments of Pb inclusions in an Al matrix.\cite{Gabrisch01} The close correspondence between these temperatures suggests that roughening and the onset of premelting approximately coincide for this interface.

%~~~~~~~~~~~~~~~
\begin{figure}[h]
\includegraphics[width= 0.9\columnwidth]{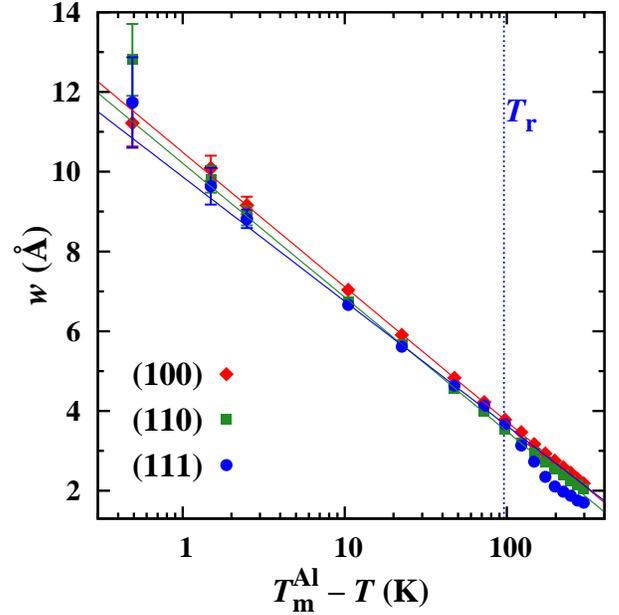}
\caption{Linear-log plot of the premelting width versus undercooling, $T_\mathrm{m}^\mathrm{Al}-T$, for each of the three interfacial orientations studied. The solid lines are the results of weighted least-squares fits to Eq.~\ref{eq:wlog}.}
\label{fig4}
\end{figure}

%~~~~~~~~~~~~~~~

As discussed earlier, theoretical considerations\cite{Kikuchi80,Lipowsky83} predict a logarithmic dependence of the premelting width on undercooling (Eq.~\ref{eq:wlog}). To  examine the validity of Eq.~\ref{eq:wlog} for the solid-liquid premelting transition in Al/Pb, we plot the calculated width of the premelting layer, $w$, as a function of the undercooling, $\Delta T = T_M - T$ on a linear-log plot in Fig.~\ref{fig4}. The data is well described by  Eq.~\ref{eq:wlog} for all three interfacial orientations for undercoolings up to 100K - deviations are seen at higher temperatures when the width approaches atomic dimensions, as expected.  Using a weighted least-squares linear regression over the temperature range 875K to 921K, we obtain estimates for $w_0$ and $T_0$. Central to the derivation of Eq.~\ref{eq:wlog} is the assumption that the interaction between the two interfaces bounding the premelted layer (the so-called "disjoining potential") is exponential and repulsive\cite{Broughton87}:
\begin{equation}
\phi_d(w) = \Delta \gamma e^{-w/w_0}
\end{equation}
where $w_0$ is the length scale of the interaction and $\Delta \gamma$ is given in Eq.~\ref{eq:delgam}. The quantity $T_0$ in Eq.~\ref{eq:wlog} is given by $T_0 = \Delta \gamma T_m/w_0 \rho L$, where $\rho$ is the number density and $L$ is the latent heat. The fitted values of $w_0$, $T_0$ and $\Delta \gamma$ are given in Table~\ref{tab:logfit}. This fitting gives an estimate of about 1.4 to 1.5 $\rm{\AA}$ for the range of the disjoining potential that is relatively independent of orientation.

%~~~~~~~~~~~~~~~ table
\begin{table}[h]
\caption{Values of $w_0$, $T_0$ and $\Delta \gamma$ (defined in Eqs.~\ref{eq:delgam} and~\ref{eq:wlog}) from a weighted linear least-squares regression of the data for $w$ versus $\ln{(T_m - T)}$. Values in parentheses represent 95\% confidence level error estimates in the last digits shown.}
\begin{tabular}{|l||l|l|l|}
\hline
&(100)&(110)&(111)\\
\hline
\hline
$w_0$ (\AA) & 1.47(3) & 1.46(3) & 1.36(4)\\
$T_0$(K) & 1.28(12)$\times10^3$ & 1.09(9)$\times10^3$ & 1.45(14)$\times 10^3$\\
$\Delta \gamma$ (mJ m$^{-2}$) & 174(16) & 148(13) & 183(18)\\
\hline
\end{tabular}
\label{tab:logfit}
\end{table}
%~~~~~~~~~~~~~~~

{\em Discussion and Summary:} Using MD simulation we predict  the existence of a solid-liquid interface premelting transition at the interface between solid Al and liquid Pb. That is, as the melting point of Al is approached from below, the surface of the crystalline Al melts to form a premelting layer of liquid Al separating the solid Al and liquid Pb bulk phases. This transition was seen in the simulations for all interfacial orientations studied - (100), (110) and (111). Although solid-vapor and grain-boundary premelting transitions are well established in the literature, premelting of a solid-liquid interface has not, to our knowledge, been previously reported at ambient pressures. Such a transition requires that the melt phase of the solid and the bulk liquid be mutually immiscible, which is true for the Al/Pb system studied here. The width of the premelting layer is shown to depend logarithmically on the undercooling $\Delta T$, as predicted by theoretical considerations.\cite{Kikuchi80,Lipowsky83,Broughton87}

At lower temperatures, near the melting point of Pb, we have previously shown that the (100) and (110) interfaces are rough, whereas the (111) interface is faceted - in agreement with experimental observations on liquid Pb inclusions in a solid Al matrix.\cite{Gabrisch01} In the current simulations, we observe a change in the activation energy of Al surface diffusion at the (111) interface at 826(4)K, which correlates well with both the experimental observations of a roughening transition at 825K for this orientation\cite{Gabrisch01} and with the observed onset of premelting in the present simulations (as evidenced by the logarithmic dependence of the premelting width on $T$).  There are a number of other solid-liquid interfacial systems in which the melt phase of the solid is immiscible in the bulk liquid, such as, for example, the interface between ice and liquid hydrocarbons, so further study of possible premelting in such systems is warranted.

%==============================================================================
\section{Acknowledgements}
%==============================================================================
YY and BBL acknowledges funding from the National Science Foundation under grant CHE-0957102. MA acknowledges support from the US National Science Foundation under Grant No. DMR-1105409.
%==============================================================================

%================================================================

%\begin{figure}[h]
%%\includegraphics[width= .90\columnwidth]{fig1.pdf}
%\end{figure}

%\begin{figure}[h]
%%\includegraphics[width= 0.9\columnwidth]{fig2.pdf}
%\end{figure}

%\begin{figure}[h]
%%\includegraphics[width= 0.9\columnwidth]{fig3.pdf}
%\end{figure}

%\begin{figure}[h]
%%\includegraphics[width= 0.9\columnwidth]{fig4.pdf}
%\end{figure}

%==============================================================================
%==============================================================================

\begin{thebibliography}{10}
\bibitem{disclaimer}
For real systems with long-range forces (e.g. dispersion), the logarithmic divergence of the width does not strictly hold in the limit that $T \rightarrow T_\mathrm{m}$, but must give way very close to $T_\mathrm{m}$ to either a power-law divergence or a finite threshold (incomplete premelting).\cite{Dash06} Such long-range forces are absent from the EAM model used here; however, the observed premelting widths do not exceed a few lattice constants and short-ranged structural forces are expected be dominant over long-range dispersion for the conditions studied.
\end{thebibliography}
\end{document}